\documentclass{aa} 
\usepackage{graphicx}
\usepackage{deluxetable}
\usepackage{txfonts}
\usepackage{natbib}
\usepackage{color} \usepackage{ulem}

\newcommand{\hii} {H\,{\sc ii}}
\newcommand{\Teff} {T$_{\rm eff}$}
\newcommand{\grav} {log\,{\em g}}

\newcommand{\ioni}[2]{{#1\,\sc{#2}}}
\newcommand{\nioni}[2]{[{#1\,\sc{#2}}]\,}

\newcommand{\Te} {$T_{\rm e}$}
\newcommand{\Ne} {$N_{\rm e}$}
\begin{document}
\title{The chemical composition of the Orion star-forming region}
\titlerunning{Stars, gas and dust in Ori\,OB1}
\subtitle{II. Stars, gas, and dust: the abundance discrepancy conundrum 
}
\author{S. Sim\'on-D\'iaz\inst{1,2} \& G. Stasi\'nska\inst{3}}
\offprints{S. Sim\'on-D\'{\i}az, \email{ssimon@iac.es}}
\institute{Instituto de Astrof\'isica de Canarias, E38200 La Laguna, Tenerife, Spain.              
           \and
           Departamento de Astrof\'isica, Universidad de La Laguna, E-38205 La Laguna, Tenerife, Spain.
           \and
           LUTH, Observatoire de Paris, CNRS, Universit\'e Paris Diderot; 5 Place Jules Janssen, 92190 Meudon, France}
\date{Received; accepted}

\abstract
{} 
{We re-examine the recombination/collisional emission line (RL/CEL) nebular abundance discrepancy 
problem in the light of recent high-quality abundance determinations in young stars in the 
Orion star-forming region. 
}  
{We re-evaluate the CEL and RL  abundances of several elements in the Orion nebula 
and  estimate  the associated uncertainties, taking into account the uncertainties in the 
ionization correction factors for unseen ions. We estimate the amount of oxygen 
trapped in dust grains for several scenarios of dust formation. We compare the 
resulting gas+dust nebular abundances with the stellar abundances of a sample of 
13 B-type stars from the Orion star-forming region (Ori\,OB1), analyzed in Papers I 
and III of this series. } 
{We find that the oxygen nebular abundance based on recombination lines agrees much better 
with the stellar abundances than the one derived from the collisionally excited lines. 
This result calls for further investigation.   If the  CEL/RL  abundance discrepancy 
were caused by temperature fluctuations in the nebula, as argued by some authors,  the same kind 
of discrepancy should be seen for the other elements, such as C, N and Ne, which is not what 
we find in the present study.  Another problem 
is that with the RL abundances, the energy balance of the Orion nebula is not well understood. 
We make some suggestions concerning the next steps to undertake to solve this problem. 
} 
{}
\maketitle
%
\section{Introduction}\label{introduction}
The knowledge of the chemical composition of stars and of the interstellar 
matter is the basis of our understanding the chemical evolution of galaxies. 
In particular, elements such as carbon, nitrogen, and oxygen provide vital clues 
for the stellar evolution and nucleosynthesis as well as on the history of 
galaxies \citep[see e.g. proceedings of CNO in the Universe by][]{Cha03}.  
\hii\ regions have for a long time been privileged sites to measure chemical 
abundances in galaxies because they were easily observable 
at large distances and because the techniques to analyze their spectra in terms of 
abundances are simple. However, the last two decades have shown the important 
potential of stars, now that computers allow the calculation of very elaborate 
stellar atmosphere models and large telescopes allow one to obtain suitable 
spectra of individual stars not only in the Milky Way, but also in other galaxies. 
Indeed, stars are superior to nebulae as abundance indicators in that 
they are not affected by depletion into dust grains.  
Another problem with \hii\ regions is that depending on the methods used 
to derive abundances, the results are different. When using collisional emission 
lines (CELs), the derived abundances are systematically smaller than when using 
recombination lines (RLs) \cite[see][and references therein]{Gar07}. 
What is the reason for these discrepancies? And which one of the two abundances 
(if any) should one believe? After several decades of studies on this, the 
debate is still not closed. \citet{Pei67} has argued for the presence of 
temperature fluctuations (of still unknown cause) that bias the nebular 
abundances derived by CELs downwards. A recent reassessment of this point of 
view can be found in \cite{Pei09}. \cite{Sta07} have examined the hypothesis 
of inhomogeneous abundances in the interstellar medium (ISM) and concluded that
if this were the case, it is the RL abundances which would be severely biased 
upwards.  
\cite{Erc09} showed that CEL-RL abundance discrepancies can be obtained 
in a chemically homogeneous medium if it contains X-ray-ionized dense clumps. 
However, she finds that the required X-ray flux to accound for the observed 
abundance discrepancy in the Orion nebula far exceeds the Orion's known stellar 
and diffuse X-ray budget. 
 
One way to attack this problem of abundance bias is to confront abundances 
derived in \hii\ regions with those derived  in B-type stars in 
the same environment, since both type of objects are expected to
share the same chemical composition, reflecting the present-day abundance pattern 
of the galactic region where they are located.
This has  been done, for example in the galaxy NGC\,300, where oxygen abundances 
have been derived for blue supergiants as well as for giant \hii\ regions by  
\cite{Bre09, Bre10}. The conclusion of that study was that stars and 
\hii\ regions (CELs) give overall the same abundances. However, in view of the 
existence of important abundance scatter in galaxies \citep[demonstrated for 
the spiral galaxy M33 by][]{Ros08}, more accurate tests are needed.  
In particular, one should compare the abundances of B-type stars with those of \textit{the associated  \hii\ region.}  

In the Milky Way, \cite{Prz08} compared the abundances of a small but representative 
sample of B-type stars in the solar vicinity with the abundances in the Orion nebula 
published by \cite{Est04} and with the abundances in the neutral ISM from absorption 
line measurements. They concluded that taking into account depletion in dust, all 
sets of abundances are remarkably similar. The abundances from \cite{Est04} were 
those derived from recombination lines, which would argue in favor of RLs giving 
the right nebular abundances. An update of the study of Przybilla et al. was however 
warranted, with a more critical analysis of CEL and RL abundances in the Orion nebula 
and a more complete stellar data base centered in the Orion star-forming region (Ori\,OB1), 
to better adress the RL/CEL abundance discrepancy.  
 
In \citet[][Paper I]{Sim10}, the first of this series of papers, we used the stellar 
atmosphere code FASTWIND \citep{Pul05} to perform a thorough self-consistent 
spectroscopic analysis with a high-quality set of spectra of 13 early B-type 
stars from the various subgroups comprising the Orion\,OB1 association. In this
study it was shown that the dispersion of O and Si abundances between stars in 
the various subgroups found in previous analyses \cite[e.g.][]{Cun92, Cun94} was 
a spurious result and the consequence of a bad characterization of the abundance 
errors propagated from the uncertainties in the stellar parameter determination.
This result is confirmed in Nieva \& Sim\'on-D\'iaz (in prep., Paper III) by means of an independent
spectroscopic analysis with the Atlas9~code \citep{Kur93}, and recent 
versions of DETAIL and SURFACE \citep{Gid81,But85}. In paper III we
also extend the elements analyzed to C, N, Mg, Ne, and Fe.

In this paper, we focus on the stellar vs. nebular abundance comparison taking 
into account the effect of depletion onto dust grain in the nebula. The paper is structured 
as follows: In Sect. 2 we present a brief summary of the abundances in B-type stars 
in the Orion OB1 association from our recent studies. In Sect. 3 we present a 
critical analysis of CEL and RL abundances of various elements in the Orion nebula. 
In Sect. 4 we analyze the depletion of the elements in the dust phase that is occuring 
in the Orion nebula. In Sect. 5 we draw inferences from the comparison of stellar 
and nebular abundances in Orion. The main conclusions of this work are summarized in 
Sect. 6. 

\section{Abundances in B-type stars}\label{bstars}

Table \ref{steab} summarizes the stellar abundances derived in Papers I and III. Two 
types of uncertainties are given along with the mean values. The first one refers to 
the dispersion of abundances obtained for the 13 analyzed stars. The second one (in 
parentheses) indicates the uncertainties intrinsic to the method of abundance determination 
in a given star (i.e. accounting for the effect of stellar parameters, microturbulence, 
and line-by-line abundance dispersion\footnote{For Mg, the corresponding value
does not account for line-by-line abundance dispersion, because only one line is available 
for the abundance analysis.}). The derived abundances agree with those computed 
by \cite{Prz08} for a small sample of B-type stars in the solar vicinity, confirming 
the homogeneity of present-day abundances in the solar neighborhood as derived by 
young massive stellar objects already suggested by these authors. While the intrinsic 
uncertainties are on the order of 0.1\,dex, the abundances obtained in the various stars 
of our sample are remarkably coherent and lead to an unprecedented accuracy of the overall 
stellar abundance pattern in Orion. 

%
\begin{table}[!t]
{\small
\begin{center}
\caption{\small Summary of abundances in B-type stars in Ori\,OB1. Data from 
\citet[][Paper I]{Sim10} and Nieva \& Sim\'on-D\'iaz (in prep., Paper III). Mean value and standard deviation of abundances
obtained for the sample of stars. Values in parentheses indicates the intrinsic 
uncertainties of the analysis of individual stars. 
\label{steab}
}
\begin{tabular}{cccc}
\noalign{\smallskip}
\tableline\tableline
    & $\epsilon$(X)     &   & $\epsilon$(X) \\
\tableline
C  & 8.35 $\pm$ 0.03 (0.09) & Si & 7.51$^\dagger$ $\pm$ 0.03 (0.08) \\
N  & 7.82 $\pm$ 0.07 (0.09) & Mg & 7.57 $\pm$ 0.06 (0.03) \\
O  & 8.74$^\dagger$ $\pm$ 0.04 (0.10) & Fe & 7.50 $\pm$ 0.04 (0.10) \\
Ne & 8.09 $\pm$ 0.05 (0.09) &    &                        \\
\tableline
\end{tabular}
\tablenotetext{\dagger}{\footnotesize Mean value from both studies.}
\end{center}
}
\end{table}

\section{Gas phase abundances in the Orion nebula\label{gasphase}}
\begin{table*}[!t]
{\small
\begin{center}
\caption{\small Newly computed densities, temperatures, and ionic abundances from optical spectra of 
the Orion nebula. Quartiles are indicated in small fonts.  The references for the corresponding 
observations are indicated in the text.
\label{oxygen-ab}
}
\begin{tabular}{lcccccc}
\noalign{\smallskip}
\tableline\tableline
Observ.  & \Ne(OII)   & \Te(OIII)   &  \Te(NII) &  $\epsilon$(O$^+$)  & $\epsilon$(O$^{2+}$)   &  $\epsilon$(O) \\
         & [10$^3$ cm$^{-3}$] & [10$^3$ K]  & [10$^3$ K] &            &                        &                \\
\tableline
\noalign{\smallskip}
E04      &  5.10$_{\ 4.72}^{\ 5.69}$  &  8.31$_{\ 8.28}^{\ 8.33}$ & 10.3$_{\ 10.1}^{\ 10.5}$ & 7.76$_{\ 7.72}^{\ 7.80}$ & 8.44$_{\ 8.43}^{\  
8.44}$ & 8.52$_{\ 8.51}^{\ 8.53}$ \\
\noalign{\smallskip}
E98 (1)  &  3.68$_{\ 2.63}^{\ 6.25}$  &  8.21$_{\ 7.71}^{\ 8.63}$ & 10.1$_{\ 8.97}^{\ 10.7}$ & 7.70$_{\ 7.57}^{\ 7.98}$ & 8.41$_{\ 8.26}^{\  
8.56}$ & 8.49$_{\ 8.37}^{\ 8.67}$ \\
\noalign{\smallskip}
E98 (2)  &  6.08$_{\ 3.83}^{\ 12.7}$  &  8.15$_{\ 7.69}^{\ 8.37}$ & 10.9$_{\ 9.01}^{\ 12.0}$ & 7.79$_{\ 7.59}^{\ 8.22}$ & 8.46$_{\ 8.35}^{\  
8.62}$ & 8.54$_{\ 8.47}^{\ 8.82}$ \\
\noalign{\smallskip}
M09 (N)  &  2.17$_{\ 2.00}^{\ 2.37}$  &  8.15$_{\ 7.95}^{\ 8.30}$ & 9.65$_{\ 9.38}^{\ 9.94}$ & 7.97$_{\ 7.92}^{\ 8.02}$ & 8.37$_{\ 8.33}^{\  
8.41}$ & 8.51$_{\ 8.48}^{\ 8.55}$ \\
\noalign{\smallskip}
M09 (HH) & 22.00$_{\ 16.7}^{\ 35.0}$  &  8.73$_{\ 8.49}^{\ 8.85}$ & 8.91$_{\ 7.99}^{\ 9.40}$ & 8.42$_{\ 8.23}^{\ 8.75}$ & 8.11$_{\ 8.08}^{\  
8.18}$ & 8.59$_{\ 8.46}^{\ 8.85}$ \\
\noalign{\smallskip}
B06 (N)  &  4.05$_{\ 3.98}^{\ 4.13}$  &  8.48$_{\ 8.46}^{\ 8.51}$ & 10.0$_{\ 9.96}^{\ 10.0}$ & 7.87$_{\ 7.86}^{\ 7.87}$ & 8.39$_{\ 8.38}^{\  
8.40}$ & 8.51$_{\ 8.50}^{\ 8.51}$ \\
\noalign{\smallskip}
B06 (HH) &  9.33$_{\ 4.36}^{\ 28.6}$  &  8.41$_{\ 8.11}^{\ 8.47}$ & 9.80$_{\ 7.49}^{\ 10.5}$ & 7.51$_{\ 7.27}^{\ 8.27}$ & 8.70$_{\ 8.68}^{\  
8.79}$ & 8.73$_{\ 8.70}^{\ 8.91}$ \\
\noalign{\smallskip}
\tableline
\end{tabular}

\end{center}
}
\end{table*}

The gas phase abundances in an ionized nebula can be readily obtained from observed 
intensity ratios of its emission lines through methods that are well-known and 
described e.g. in \cite{Ost06}. After correcting the observed line fluxes for 
interstellar extinction, determining the temperature and density of the emitting 
zones with appropriate line ratios, the ionic abundances are obtained directly 
from emission line ratios. Elemental abundances are then obtained by summing the 
abundances of the observed ions and accounting for unseen ionic species with 
ionization correction factors (icfs). The chemical composition determined in this 
way depends on the quality of the observational data, the adopted dereddening 
procedure, the temperature, density indicators, and the chosen icfs. 
The procedures are quite standard, but may deviate in some detail from one author 
to another. Of course, the results depend on the set of atomic data that are used. 

\subsection{O}
\label{O}
  
\paragraph{O from CELs}
\label{OfromCELs}
 
Oxygen is the element for wich the abundances can be determined most reliably in 
\hii\ regions, because both ions that are expected to be present, O$^+$ and O$^{2+}$, 
emit strong lines in the optical and, in addition, the temperature and density of 
the emitting zones can be easily obtained from appropriate line ratios. 

There have recently been a number of optical observations of the Orion nebula with 
high resolution and very good signal-to-noise ratio. To obtain a set of abundances 
in the Orion nebula that gives a fair account of uncertainties and possible variations 
from one point to another, we rederived in a consistent way the CEL abundances 
of oxygen in different positions of the Orion nebula from the reddening-corrected 
intensities of by \citet[][E98]{Est98}, \citet[][E04]{Est04}, \citet[][B06]{Bla06}, 
\citet[][M09]{Mes09}.  We included the two slit positions provided
by E98, B06, and M09. In the case of E98, both spectra correspond to the nebula, while for
B06 and M09, one of them is located in the shock front of one of the Herbig-Haro objects 
of the Orion nebula. For the abundance analysis, we assumed that the electron 
density throughout is the one given by the [\ioni{O}{ii}]3726/3729 line ratio\footnote{Densities 
given by [\ioni{Cl}{iii}] or [\ioni{Ar}{iv}] doublets would, in principle, be more 
appropriate for the high-excitation zone, but they are much more uncertain in 
the Orion nebula. Fortunately the abundance analysis is not strongly affected by the 
density in the expected range of densities. Densities derived from the 
[\ioni{S}{ii}]6717/6731 line ratio are slightly lower than those derived from 
[\ioni{O}{ii}]3726/3729.}. We have considered that H$\beta$ and [\ioni{O}{iii}]5007 are 
emitted in a zone whose temperature is obtained from the [\ioni{O}{iii}]4363/5007 line 
ratio and that lines from [\ioni{O}{ii}]3727 and [\ioni{N}{ii}]6584 are emitted in a 
zone whose temperature is obtained from the [\ioni{N}{ii}]5755/6584 line ratio. The 
references for the atomic data used in the calculations are listed in Table 
\ref{atomdat} in  Appendix A. The intensities of the forbidden lines were obtained from a 5-level 
model atom.

The resulting electron densities, temperatures, and abundances of  O$^+$, O$^{2+}$, and O 
in units of $\epsilon$(X)= 12 + log X/H are listed in Table \ref{oxygen-ab}. The 
uncertainties (listed as quartiles in the table) were obtained by Monte-Carlo 
simulations, taking into account observational errors in the dereddened line intensities 
as listed in the original papers\footnote{We have taken into account the fact that 
electron densities are obtained from adjacent lines whose measured intensity ratios do 
not depend on calibration errors, but only on line-fitting and signal-to-noise ratios. }. 

We find that  CEL results are remarkably consistent with each other once the same 
atomic data and procedures are applied to all the spectra,  when disregarding 
the two Herbig-Haro objects (labelled HH in the Table).  This is because of the exceptional 
quality of the data reported in the original papers. As for our B stars, the 
dispersion of the derived CEL oxygen abundances is smaller than the formal errors:  
excluding the two Herbig-Haro object, the average value of  $\epsilon$(O) is  8.51 and 
the dispersion is 0.02\,dex. For convenience, below we will exclusively use 
the abundances derived from the spectra by E04. 
\paragraph{O from RLs}
\label{OfromRLs}

The method using recombination lines to determine ionic abundances has 
the advantage of being almost independent of errors in the electron temperature. On the 
other hand, it relies on very weak lines and requires a certain know-how to choose the 
best strategy for the different observed multiplets. We therefore adopted the ionic abundances 
derived in the original paper. Note that the abundances derived by E04 
from individual recombination lines of the same ion differ much more than expected 
from observational errors only. The final adopted abundances are a weighted mean of 
the sum-values of the most reliable multiplets. The error attributed to the final 
RL abundances by E04, which we adopt here as well, is also a result of this weighting 
process. 

\cite{Est04} derive RL abundances for both O$^+$ and O$^{2+}$. But they note that the RL  
O$^+$ abundance, which is based on a single faint line in a spectral region much affected by 
sky emission,  is not very reliable. We will thus use only their abundance of O$^{2+}$, 
and from it derive the RL oxygen abundance using the O$^{2+}$/O$^{+}$ obtained from 
forbidden lines. The result is presented in Table \ref{table-RLocne}.

\begin{table*}[!t]
{\small
\begin{center}
\caption{\small RL abundances of O, C, and Ne from the E04 optical spectrum of 
the Orion nebula. 
\label{table-RLocne}
}
\begin{tabular}{cccc}
\noalign{\smallskip}
\tableline\tableline
 & O$_{\rm RL}$  & C$_{\rm RL}$ & Ne$_{\rm RL}$\\
\tableline
\noalign{\smallskip}
ion abundance  & 
$\epsilon$(O$^{2+}$)$=$8.57$\pm$0.01 & 
$\epsilon$(C$^{2+}$)$=$8.34$\pm$0.02 & 
$\epsilon$(Ne$^{2+}$)$=$7.95$\pm$0.09  \\

icf & 
log icf (O$^{2+}$)$=$0.08$\pm$0.02  & 
log icf (C$^{2+}$)$=$0.03$\pm$0.01 & 
log icf (Ne$^{2+}$)$=$0.29$\pm$0.02 \\
element abundance  &
$\epsilon$(O)$=$8.65$\pm$0.03 & 
$\epsilon$(C)$=$8.37$\pm$0.03 & 
$\epsilon$(Ne)$=$8.24$\pm$0.11\\
\tableline
\end{tabular}
\end{center}
}
\end{table*}

\subsection{N, Ne, S, Ar}
\label{nnesar}

\paragraph{N, Ne, S, Ar from CELs}
\label{nnesarCELs}

\begin{table*}[!t]
{\small
\begin{center}
\caption{\small Newly computed CEL abundances of N, Ne, Ar, S from the E04 optical spectrum of 
the Orion nebula. 
\label{table-nnears}
}
\begin{tabular}{ccccc}
\noalign{\smallskip}
\tableline\tableline
  & N$_{\rm CEL}$  & Ne$_{\rm CEL}$ & Ar$_{\rm CEL}$ & S$_{\rm CEL}$ \\
\tableline
\noalign{\smallskip}
ion abundance  & 
$\epsilon$(N$^{+}$)$=$6.92$\pm$0.05 & 
$\epsilon$(Ne$^{2+}$)$=$7.76$\pm$0.01  &
$\epsilon$(Ar$^{2+}$)$=$6.37$\pm$0.02  &
$\epsilon$(S$^{2+}$)$=$6.80$\pm$0.05 \\
icf  & 
log icf(N$^{+}$)$=$1.00$\pm$0.04 & 
log icf (Ne$^{2+}$)$=$0.29$\pm$0.02 &
log icf (Ar$^{2+}$)$=$0.02$\pm$0.01 &
log icf (S$^{2+}$)$=$0.07$\pm$0.01 \\
element abundance  & 
$\epsilon$(N)$=$7.92$\pm$0.09 & 
$\epsilon$(Ne)$=$8.05$\pm$0.03 &
$\epsilon$(Ar)$=$6.39$\pm$0.03 &
$\epsilon$(S)$=$6.87$\pm$0.06 \\
\hline
\end{tabular}
\end{center}
}
\end{table*}

We use the same procedure as for oxygen to derive the CEL ionic abundances of 
N$^{+}$, Ne$^{2+}$, Ar$^{2+}$, S$^{+}$, and S$^{2+}$ from the E04 spectrum, 
using the atomic data listed in Table \ref{atomdat}. Like for oxygen, the 
uncertainties in the ionic abundances are determined from Monte-Carlo simulations 
of the observational errors. The resulting ionic abundances are displayed in 
Table \ref{table-nnears}. To obtain the corresponding elemental abundances, we need 
to use ionization correction factors. Previous studies (E98, E04) used icfs from a 
variety of sources that are not necessarily consistent and sometimes outdated. To 
obtain the most reliable set of icfs for our purpose, we constructed a 
tailored photoionization model of the Orion nebula with the Cloudy code \citep{Fer98}. 
The way we proceeded to 
obtain the model is described in detail in Appendix \ref{model}. From this model we derive 
icfs for N, Ne, S and Ar\footnote{Ar and S abundances have  not been determined 
so far in our sample of stars, so they will not be used in this study. However, the 
stellar values should be available in the near future.}. 
Those and the resulting elemental abundances are listed 
in Table \ref{table-nnears}. There are non negligible 
differences between our total abundances and the ones found by E04 \citep[and][]{Gar07}.  
They stem essentially from the different icfs adopted in each case. 
Our values should be more reliable, because they were obtained from a 
tailored model. Note that some icfs are large, especially that of N$^{+}$, which we 
find to be on the order of 10 (E04 attributed a value of about 6). In this case, even 
with a tailored photoionization model, we cannot be entirely confident in the result 
we find. Concerning neon, the icf derived from our model implies that Ne/O is 
not equal to Ne$^{++}$/O$^{++}$, which is the classical recipe to derive the neon abundance. 
Using the spectroscopic data for  about 80 zones in the Orion nebula from \citet{Ode10},  
we find that  Ne$^{++}$/O$^{++}$ shows no trend with O$^{++}$/H over a range of one 
decade in  O$^{++}$/H. This would rather support the classical icf for Ne over the one we 
derived from our model. Indeed, model atmospheres become more uncertain at 
energies above 40 eV, and the nebular constraint provided by our photoionization model is 
given by the very weak \nioni{Ar}{iv} lines. On the other hand, \citet{Rub10} used 
infrared line intensities measured by the \textit{Spitzer} Space Telescope to derive  
12 + log (Ne/H) = 8.00 $\pm$ 0.03  without the need of any icf because both  
\nioni{Ne}{ii} and  \nioni{Ne}{iii} lines are observed. This value is very close to the 
one we derive with our model icf. In view of this dilemma, we give in Table \ref{sum-ab} 
the values of Ne derived both with our model icf and with the classical icf.


\paragraph{Ne from RLs}
\label{nnesarRLs}

Of the elements N, Ne, S and Ar, reliable abundances from recombination lines
 exist only for Ne, using the Ne$^{2+}$ ion. The results are listed in 
Table \ref{table-RLocne}, again using both the model icf and the classical one.

\subsection{C}
\label{C}

For carbon there are no CELs in the optical wavelength range, and one has 
to resort to ultraviolet observations. There are several difficulties 
with this: the extinction correction is important and may be inaccurate. 
Collisionally excited ultraviolet lines depend very strongly on the electron 
temperature, so using temperatures derived from [\ion{N}{ii}] or [\ion{O}{iii}] 
lines may lead to substantial errors. \cite{Rub93} performed a detailed 
photoionization modeling to infer the carbon abundance by direct comparison of 
model results with observational data in several locations of the Orion nebula. 
They point out however that their model was not able to explain  
the \ioni{C}{iii}]\,1907,09 and the \ioni{C}{ii}]\,2324-29 lines at the same time. As a result, 
their carbon abundance is not very accurate. We will use the value 
$\epsilon$(C) = 8.4 and assign to it an uncertainty of $\pm 0.25$\,dex	 
 
In the optical, there are several RLs that allow one to determine the abundance 
of C$^{2+}$. We use the value determined by E04 and list it in Table \ref{table-RLocne}.  
To determine the total carbon abundance from RLs, we use the icf derived from our 
photoionization model, as done for the other elements. The result is listed in  
Table \ref{table-RLocne}.
 
\subsection{Fe}
\label{Fe}
 
There are many forbidden lines of Fe$^+$, Fe$^{2+}$ and Fe$^{3+}$ in the optical 
spectra of the Orion nebula. To make the best use of them, multilevel atoms of 
over 15 levels have to be considered. \cite{Rod05} performed a detailed analysis 
of the Fe spectrum in the Orion nebula using the spectrum of E04, and compared 
the Fe abundance obtained by summing the observed abundances of Fe$^+$, Fe$^{2+}$ 
and Fe$^{3+}$ with the abundance derived using an icf implied by their 
photoionization model analysis. They found an inconsistency of about a factor 4, 
implying that one or several of the atomic data involved in the analysis are wrong.  
  
\subsection{Mg and Si}
\label{mgandsi}
 
Collisionally excited lines of Si and Mg  ions are not present in the optical range. 
Observations are available in the UV:  \ioni{Si}{iii}]\,1883,93 and \ioni{Mg}{ii}\,2768. 
The former have been analyzed by \cite{Rub93}, together with \ioni{C}{iii}]\,1907,09. 
These authors find a Si/C ratio of 0.016 with an accuracy of about 30\%. The absolute 
abundance of Si is plagued however with the uncertainty in the C abundance as deduced 
from UV lines. The abundance of Mg was inferred from a photoionization modeling by 
\cite{Bal91}, but its value is not accurate, because the observed 
intensity of \ioni{Mg}{ii}\,2768 is very uncertain. In addition, the flux of \ioni{Mg}{ii}\,2768 
received at Earth is affected by interstellar absorption. 

\begin{table*}[!ht]
{\small
\begin{center}
\caption{\small Summary of gas-phase abundances in the Orion nebula.
\label{sum-ab}
}
\begin{tabular}{llllll}
\noalign{\smallskip}
\tableline\tableline
A  & $\epsilon$(A)	 	& 	data  	& 	 source  		& method  & major uncertainties\\
\tableline
\noalign{\smallskip}
C  & 8.40$\pm$0.25  & CEL: \ioni{C}{iii}]1907,09 & \cite{Rub93}		&	 model & dereddening, CII  not fitted	 \\  
   & 8.37$\pm$0.03  & RL:  \ioni{C}{ii} 	&    C$^{++}$: E04          &	icf (own model)			&  \\
\noalign{\smallskip}   
N  & 7.92$\pm$0.09  & CEL: [\ioni{N}{ii}]6584 &   data: E04, N$^{+}$: this paper             &   icf (own  model)   &	icf		 \\  
\noalign{\smallskip}
O  & 8.52 $\pm$0.01 & CEL: [\ioni{O}{iii}]5007, [\ioni{O}{ii}]3727 &  data: E04, O$^{+}$ \& O$^{++}$: this paper            	&  no icf needed &  \\ 
   & 8.65$\pm$0.03  & RL:  \ioni{O}{iii}      &   O$^{++}$: E04        &		 icf (own  model)		&   \\  
\noalign{\smallskip}
Ne & 8.05$\pm$0.03  & CEL: [\ioni{Ne}{iii}]3869  &  data: E04,  Ne$^{++}$: this paper  & icf (own model)  & \\
& 7.84$\pm$0.03  & CEL: [\ioni{Ne}{iii}]3869  &  data: E04,  Ne$^{++}$: this paper  & icf (classical)  & \\
\noalign{\smallskip}
   & 8.25$\pm$0.35  & RL: \ioni{Ne}{ii}   &    Ne$^{++}$: E04$^a$  & icf (own model)  & \\
   & 8.03$\pm$0.26  & RL: \ioni{Ne}{ii}   &    Ne$^{++}$: E04$^a$  & icf (classical)  & \\
\noalign{\smallskip}
S  & 6.87$\pm$0.06  & CEL: [\ioni{S}{iii}]9069  &  data: E04,  S$^{++}$: this paper  & icf (own model)  & \\
\noalign{\smallskip}
Ar & 6.39$\pm$0.03  & CEL: [\ioni{Ar}{iii}]7135  &  data: E04,  S$^{++}$: this paper  & icf (own model)  & \\
\noalign{\smallskip}
Mg & 6.50 :: 	    & CEL: \ioni{Mg}{ii} 2798 & \cite{Bal91}    &	 model				& intensity, interstellar absorption\\  
\noalign{\smallskip}
Si & 6.50$\pm$0.25  & CEL: \ioni{Si}{iii}]1883,92  &  \cite{Rub93}	     & model			& Si/C more accurate than Si/H\\  
\noalign{\smallskip}
Fe & 6.0$\pm$0.3    & CEL: [\ioni{Fe}{ii}], [\ioni{Fe}{iii}], [\ioni{Fe}{iv}] &   \cite{Rod05}    & 	model 	& sum of observed ions gives \\
 & & & & & different  result\\  
\noalign{\smallskip}
\tableline
\end{tabular}
\tablenotetext{a}{\footnotesize The uncertainty for Ne$^{++}$ was erroneaously given as 0.09 dex in E04, instead of 0.2 dex (J. Garc\'{\i}a-Rojas, private communication).}
\end{center}
}
\end{table*}

 \subsection{Summary of gas phase abundances}
 
In Table \ref{sum-ab} we summarize our nebular abundance analysis by giving our adopted 
values of the gas phase C, N, O, Ne, S, Ar, Mg, Si, and Fe abundances. The error bars  
now include the uncertainties from the Monte-Carlo analysis and the uncertainties in the 
icfs, all added in quadrature. The element with the best determination is oxygen. For 
nitrogen, the large icf introduces an uncertainty difficult to evaluate, but possibly 
larger than indicated in the table. For neon, the icf derived from our photoionization 
model leads to an abundance larger by 0.2  than the one obtained using a classical icf. 
This issue needs to be investigated further. The abundance of carbon from CELs is quite uncertain, 
as are the abundances of the refractory elements Fe, Si and Mg, each one for different  reasons.
 
\begin{figure}[t!]
   \centering
   \includegraphics[width=7.5cm, angle=90]{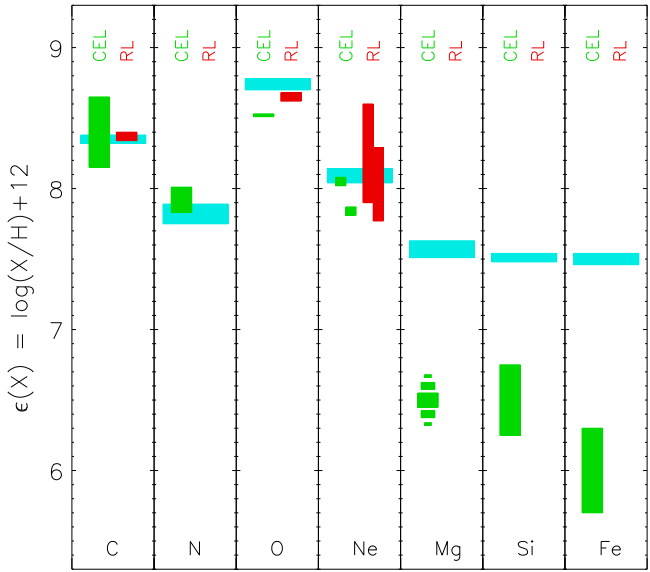}
      \caption{Comparison of C, N, O, Ne, Mg, Si, and Fe abundances obtained 
      from B-type stars (cyan), CEL (green), and RL (red). For Ne there are 
      two green boxes, corresponfing to different icfs (see text). The height of 
      the boxes represents the uncertainties, indicated in Tables \ref{steab} and \ref{sum-ab}.
      Note that for Mg there is no evaluation of the uncertainty.
      See text for more detail.}
         \label{fig2}
   \end{figure}

Figure \ref{fig2} compares the abundances of C, N, O, Ne, Mg, Si, and Fe in the Orion nebula obtained 
using CELs (green boxes) and RLs (red boxes) with the abundances in the B stars of  Orion\,OB1 (cyan boxes). 
In each case, the height of the boxes represents the  uncertainties, indicated in Table \ref{steab} (standard
deviation from the 13 stars) and Table \ref{sum-ab} (Orion nebula).
 
\section{Depletion of the elements in the dust phase}\label{sec-dust}

Before comparing the nebular abundances with the stellar ones, one has to consider
that some of the material in the nebula may be in the form of dust 
grains, whose total mass and chemical composition cannot be directly accessed. That 
the abundances of the refractory elements Mg, Si, and Fe in the gas phase 
of the Orion nebula (Table \ref{sum-ab}) are lower than those found in the 
atmospheres of our sample of B stars (Table \ref{steab}) by at least one order of magnitude 
 clearly indicates that these elements are mostly found in 
dust grains. We find, by comparing the numbers in those two tables, that 
 $\sim$\,90\% of Mg, $\sim$\,80\,--\,90\% of Si, and $\sim$\,93\,--\,98\% 
of Fe are trapped in dust grains. If we knew the chemical composition of dust, we 
could estimate the degree of depletion of the other elements.

According to \cite{Dra03}, silicates of the type Mg-Fe pyroxenes
(Mg$_{\rm x}$Fe$_{\rm 1-x}$SiO$_3$) and olivines (Mg$_{\rm 2x}$Fe$_{\rm 2-2x}$SiO$_4$)
contribute a substantial fraction of the total mass of interstellar dust. These 
molecules are the main cause for the depletion of Si. In addition, Mg, Fe, and O could 
be in oxide form (MgO, FeO, Fe$_2$O$_3$, Fe$_3$O$_4$). To estimate the amount of oxygen 
depletion in the Orion nebula, we investigate all combinations of dust grain 
compositions presented in Table \ref{dustcomp}.  We implemented and IDL procedure 
that calculates O$_{\rm dust}$ for each of the combinations. The input parameters
of the program are the stellar and nebular Si, Fe, and Mg abundances, plus the 
nebular gas phase oxygen abundance. Once the total amount of Si, Fe, and Mg available to form 
dust grains is computed, the program determines the amount of O, Fe, and Mg used
to form silicates (the whole amount of Si depleted is used to this aim); then, with 
the remaining Fe and Mg available, the amount of O used to form Fe and Mg oxides 
is derived. Finally, the total amount of oxygen in the nebula (gas + dust) is determined 
for each combination of dust grain composition by adding the amount of oxygen
depleted in silicates and oxides to the nebular gas phase abundance.

%
\begin{table}[!t]
{\small
\begin{center}
\caption{\small Dust composition possibilities considered in this study. 
Values in parentheses correspond to the fraction of molecules from each type. 
\label{dustcomp} }
\begin{tabular}{cc}
\noalign{\smallskip}
\tableline\tableline
Silicates & Composition \\
\tableline
A & (x)MgSiO$_3$, (1-x)FeSiO$_3$ \\
B & MgFeSiO$_4$\\
C & (x)Mg$_2$SiO$_4$, (1-x)Fe$_2$SiO$_4$ \\
Others & A+B, A+C, B+C, A+B+C \\
\noalign{\smallskip}
\tableline
\noalign{\smallskip}
Mg oxides & Composition \\
\tableline
\noalign{\smallskip}
 & MgO\\
\noalign{\smallskip}
\tableline
\noalign{\smallskip}
Fe oxides & Composition \\
\tableline
\noalign{\smallskip}
 & (x)FeO, (y)Fe$_2$O$_3$, (1-x-y)Fe$_3$O$_4$\\
\noalign{\smallskip}
\tableline
\end{tabular}
\end{center}
}\end{table}

\begin{figure}[t!]
   \centering
   \includegraphics[width=9.5cm, angle=90]{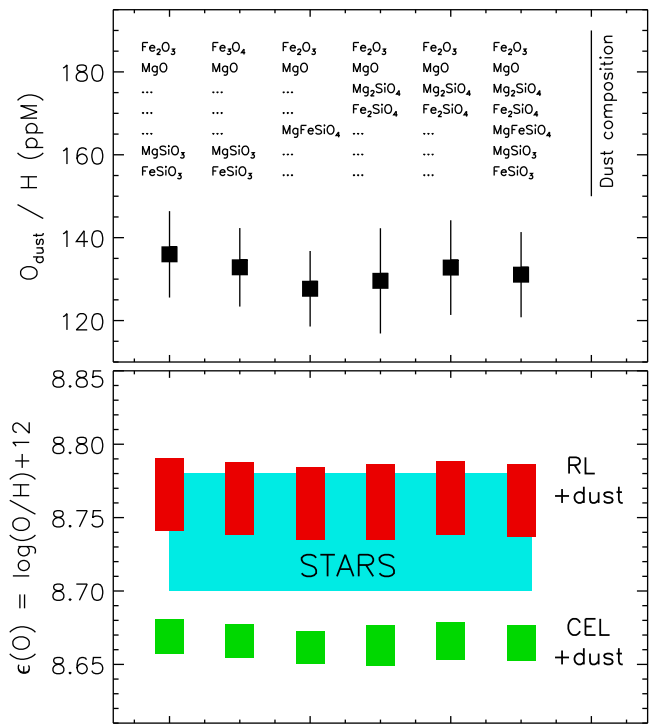}
      \caption{Upper panel: Amount of oxygen expected to be in solid form for 
          various dust composition scenarios (see Table \ref{dustcomp}).
	  Lower panel: Comparison of the stellar oxygen abundance
	  (wide cyan rectangle) with the gas+dust oxygen abundance in the nebula  
          for the different dust scenarios. Green rectangles are obtained with 
          CEL gas abundances, red rectangles are obtained with RL gas abundances. 
          See text for an explanation of the considered uncertainties.}
         \label{oxygendepl}
   \end{figure}

We ran the IDL program 1000 times with random independent values for each of the input 
abundances\footnote{Each of the input abundances has a normal distribution with mean 
values and standard deviations as those indicated in Tables \ref{steab} and \ref{sum-ab}.} 
to derive mean values and uncertainties of the amount of oxygen trapped in dust for each 
of the dust grain compositions considered. The results are presented in the upper panel of 
Fig. \ref{oxygendepl}. We find that the oxygen depletion is  in the range
125\,-\,135 ppM, with corresponding uncertainties of $\sim$\,10 ppM. This agrees 
very well with the value of 0.12 dex estimated by \citet{Mes09}.

The total amount of oxygen in the nebula (gas + dust) determined for each 
combination of dust grain composition is presented in the bottom panel 
of Fig. \ref{oxygendepl}. Two cases are shown: the green boxes represent the 
results using CEL abundances, while the red boxes represent the results using 
RL abundances. We see that the correction 
for O depletion owing to dust grains does not depend much on the details of the dust 
composition in the different combinations of dust grains we considered\footnote{For 
completeness, we mention that \cite{Jen09} inferred from a detailed study of 
absorption line measurements in the solar vicinity that a significant 
proportion of interstellar oxygen should be trapped in ices. This statement, however, 
does not affect our estimates, since ices are not expected to survive in an ionized 
nebula, because of the effect of photosputtering \citep{Gri07}.}.
For comparison with the total nebular value, the stellar oxygen abundance is 
given in cyan. The height of the green and red  boxes represent the error 
bars combining the uncertainties in the gas phase abundance determination and in 
the dust phase abundance. The height of the cyan box represents the dispersion in 
the oxygen stellar abundances given in Table \ref{steab}. 
The verdict from Fig. \ref{oxygendepl} is clear. Oxygen abundances in the Orion 
nebula derived from RLs, when corrected for the effects of depletion in dust grains, are very similar 
to those found in the B stars of the Ori\,OB1 association. This is not the case for the 
oxygen CEL abundances, which lead to much lower values. We note
that the oxygen CEL abundances would actually be compatible with the stellar abundances 
if using the intrinsic uncertainties of the analysis of individual stars (see Table \ref{steab}). 
But as argued in Sect. \ref{bstars}, the degree of coherence between the abundances 
obtained for the 13 studied stars indicates that the oxygen abundance of the Ori\,OB1 cluster 
is determined with much better accuracy than $\sim$\,0.10 dex.
  
Concerning the other elements considered in this study, there are observational 
arguments that a significant fraction of carbon atoms must be locked in carbon-based 
dust and in PAHs although the amount of carbon depletion in nebulae is still very poorly 
known \citep{Dra03, Jen09}. To our knowledge, there is no indication that nitrogen 
could be present in solid form, except perhaps in ices \citep{Hil10}, which are 
not expected to be present in ionized gas. As for neon, which is a noble gas, it is 
not expected to be incorporated in dust grains, at least not in significant amounts.

\section{Discussion}
\label{discussion}

 Our study therefore leads us to the conclusion that 
the oxygen abundances derived in the Orion nebula from CELs are incompatible
with the abundances derived in the B stars (once the effects of depletion in dust grains 
are taken into account), while RL oxygen abundances agree better (although
the mean values seem to be somewhat higher that the stellar one).

That RL abundances are more reliable than CEL abundances has been argued in many papers by 
Peimbert and coworkers \citep[see e.g.][]{Pei09}, including E04. One of the 
arguments is that the abundance discrepancy factor between RL and CEL abundances can be 
explained by temperature fluctuations of the same order as the difference between the 
temperature derived from the  \nioni{O}{iii} 4363/5007 line ratio and the one derived from 
the Balmer jump. In addition, temperature fluctuations of that order in the Orion nebula 
have been inferred from the spatial variation of \Te\nioni{O}{iii} across the nebula 
\citep{Ode03, Rub93}. 

There are, however, a certain number of controversial points to consider. 

First, when looking at Fig. \ref{fig2}, we see that the carbon RL abundance is 
slightly higher than the carbon abundance in the B stars, while we applied 
no correction for the depletion of carbon in dust grains. If we adopt the Si, Mg, and Fe 
depletions inferred from this figure, we infer a total dust-to-gas mass ratio in 
the silicates and oxides of 0.41$\pm$0.03\%. With the ``Orion mixture'' considered 
in Cloudy, this implies a carbon depletion of $\sim$\,50\%. Then the total gas+dust 
carbon abundance, as derived from RLs, would be higher than the stellar one by about 0.2 dex. 
Note that the CEL carbon 
abundance is also higher than in the stars. 
However, in that case, the error bar is large and the CEL value is not really reliable. 
The nitrogen CEL abundance is slightly above the stellar one. However, taking into 
account the large icf (see Table \ref{table-nnears}) and the larger error bar on 
stellar N abundances (see Table \ref{steab}), both values are perhaps compatible. 
For neon, the RL abundance is somewhat uncertain and does not provide a useful constraint, 
but the CEL abundance is compatible with the stellar value if taking the model icfs and 
lower by about 0.2 dex if using the classical icfs. If the oxygen RL/CEL 
abundance discrepancy is caused by temperature fluctuations, one would expect the CEL 
abundances for C, N, and Ne to be significantly below the stellar values, and the RL abundances equal to the stellar ones. 
This is clearly not what occurs for carbon (for neon, the diagnostic depends on what icf is used).

The second problem is related to the energy budget. To show this, we constructed 
a photoionization model with exactly the same parameters as the model shown in  
Fig. \ref{figmodelCEL}, except that for C, O, and Ne we take the RL abundances listed in 
Table \ref{table-RLocne} (and for N we also assume an RL-like abundance, by adding 
0.14\,dex to the CEL value). This model is shown in Fig. \ref{figmodelRL}. As seen 
in that figure, the value of \Te\nioni{O}{iii} returned by the model is below the observed 
one by 2 sigmas. In order for the model to reproduce the observed \Te\nioni{O}{iii}, 
one would need to consider extra heating, at least in some zones. Then, the computed 
intensities of the forbidden lines would be signficantly higher than observed. A model 
with slightly lower input abundances, in which the chemical composition of the nebular 
gas is \textit{exactly }the same as found for the stars (but taking depletion into account)  
gives basically the same result: the nebular model becomes too cool and additional 
heating would make the mismatch between predicted and observed forbidden line intensities 
even worse.  

Could it be, then, that the solution lies in oxygen-rich condensations, as studied by 
\cite{Sta07}? This does not seem to be the case, at least in Orion. 
As found by those authors, oxygen rich condensations would bias the RL abundances upwards , 
while we found in this paper that the RL abundances are compatible with the stellar 
ones.

\section{Conclusion}
\label{conclusion}

The aim of the present paper was to re-examine the RL-CEL nebular abundance discrepancy 
problem in the light of recent high-quality abundance determinations in young stars. In 
this study, we focused on one object: the  Orion star-forming region. There is no 
\hii\ region in the solar neigborhood in which the stellar and gas phase abundances 
can be derived with better accuracy, and the stars  from the Orion\,OB1 association, which  
share the same composition as the associated  \hii\ region, have been studied in great detail.

We have re-evaluated the CEL and RL  abundances of several elements in the Orion nebula 
and estimated the associated uncertainties, taking into account the uncertainties in the 
ionization correction factors for unseen ions. We estimated the amount of oxygen 
trapped in dust grains for several scenarios for dust formation. We compared the 
resulting gas+dust nebular abundances with the stellar abundances of a sample of 13 B-type 
stars from the Orion star-forming region (Ori\,OB1), analyzed in Papers I and III of this series. 

We find that the oxygen nebular abundance based on recombination lines agrees much better 
with the stellar abundances than that derived from the collisionally excited lines. 

However, there remain problems to be solved. First, the abundances of the other elements, C, N, and Ne, 
although admittedly less accurate, do not deliver such a clear-cut picture. If the RL-CEL 
abundance discrepancy were caused by temperature fluctuations, as argued by \cite{Pei09}, 
one should observe the same kind of bias in the CEL abundances of the other elements. 
Hopefully, the consideration of further elements, such as S and Ar, whose abundances will 
soon be available for the stars of the Orion\,OB1 association, will bring useful additional 
constraints. Another problem is that with the RL abundances, the energy balance of the Orion 
nebula is not well understood. Investigating this problem will require detailed photoionization 
modeling and an accurate account of all heating and cooling processes taking place in 
this specific case, as well as the effect of uncertainties in the collision 
strenghts for the transitions of interest. In parallel to such studies, another 
avenue to start exploring is the possibility of remaining biases in the stellar 
abundance determinations, for example because of a slight offset in the  description 
of the ionization of oxygen in the model atmospheres of B-type stars, which 
unfortunately cannot be checked directly here, since only \ioni{O}{ii} lines 
are seen in the studied stars \citep[see][]{Sim10}.

An important step forward would be  to repeat the analysis performed in this paper for 
objects where the RL-CEL discrepancy is  larger. The choice of the Orion nebula, while imposed 
by the quality of relevant observations and data 
analysis, is indeed not the best one to investigate the  RL-CEL nebular abundance 
discrepancy problem in \hii\ regions because the discrepancy there is actually quite modest: 
$\simeq$ 0.12\,dex.  With an adequate observing strategy and data handling, 
it should be possible to investigate the RL-CEL abundance discrepancy in different environments, 
for example at low metallicities. This is a necessary condition to get a definite handle on 
this exasperating problem.

\begin{acknowledgements}

Financial support from the Spanish Ministerio de Ciencia e Innovaci\'on under
the project AYA2008-06166-C03-01 is acknowledged.  This work has also been
partially funded by the Spanish MICINN under the Consolider-Ingenio 2010
Program grant CSD2006-00070: First Science with the GTC
(http://www.iac.es/consolider-ingenio-gtc).
\end{acknowledgements}

\begin{table*}[!t]
\begin{center}
\caption{\small Atomic data for line emission used in Sect. \ref{gasphase}.
\label{atomdat} }
\begin{tabular}{lll}
\noalign{\smallskip}
\tableline\tableline
ion    &  collision strengths   &  transition probabilities    \\
\hline
\noalign{\smallskip}
N II   &  \cite{Hud04}  + priv.  & \cite{Gal97}  \\ 
       &                         & \cite{Sto00}  \\       
O II   &  \cite{Tay07}  	 & \cite{Tac02}  \\
O III  &  \cite{Agg99}           & \cite{Gal97}  \\
       &                         & \cite{Men99}  \\
       &                         & \cite{Sto00}  \\ 
Ne III &  \cite{McL00}  	 & \cite{Gal97}  \\
       &                         & \cite{Sto00}  \\       
S II   &  \cite{Kee96}  	 & \cite{Fro04a} \\
S III  &  \cite{Tay99}  	 & \cite{Fro04b}  \\
Ar III &  \cite{Gal98b}  	 & \cite{Men82a}   \\
Ar IV  &  \cite{Ram97} + priv.   & \cite{Men82b}   \\
\noalign{\smallskip}
\tableline
\end{tabular}
\end{center}
\end{table*}


\appendix

\section{Atomic data for nebular emission line}
\label{model}

\section{Photoionization model}
\label{model}

To estimate the ionization correction factors as well as possible for the 
elements whose ionic abundances were derived from the optical spectrum of 
E04, we constructed a photoionization model reproducing the observed 
characteristics of this spectrum.

The model was constructed with the photoionization code Cloudy \citep[as 
last described in][]{Fer98}, using version 8.0. We considered 
an open geometry (as appropriate for the Orion nebula) and taken the 
``intrinsic line fluxes''  and ``aperture slit'' options. Of the options 
offered by Cloudy, this one seems the most approriate for the E04 observation.
 
Because we are here mainly interested in the icfs, we do not try to fit 
all observed line intensities (we already know that this would not be 
possible for RLs and CELs simultaneously). On the other hand, we aim at 
reproducing the line ratios indicative of density (\nioni{Ar}{iv}\,4740/4711,  
\nioni{O}{ii}\,3726/3729, and  \nioni{S}{ii}\,6731/6716) and ionization 
structure (\nioni{O}{iii}\,5007/\nioni{O}{ii}\,3727, \nioni{S}{iii}\,9069/(\nioni{S}{ii}\,6731+16)  
and \nioni{Ar}{iv}\,4711+40/\nioni{Ar}{iii}\,7135).
    
We assumed an exponential density law, which is usually considered for the Orion 
nebula \cite[see e.g.][]{Rub93}: $N$(H)\,=\,$N_{\rm max}$\,exp[($x$-$x_{\rm p}$)/$h$], 
with $N$(H)\,=\,$N_{\rm max}$ for $x\,>\,x_{\rm p}$. We considered as free 
parameters $h$, $N_{\rm max}$, and $x_{\rm p}$, and fixed the internal radius of
the model to be $r_0$\,=\,0.9\,$x_{\rm p}$.

\begin{figure}[t!]
   \centering
   \includegraphics[width=6cm, angle=90]{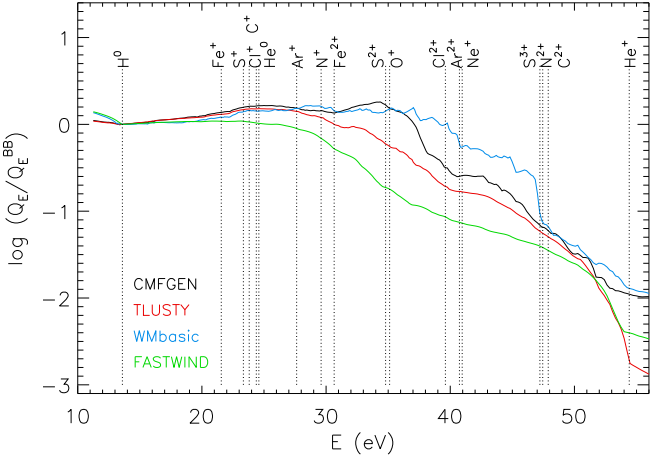}
      \caption{ Comparison of the ionizing SEDs for $\Theta^1$\,Ori\,C as 
               predicted by various stellar atmosphere models. Stellar
               parameters: \Teff=39000, \grav=4.0 \cite[derived by][by means 
               of a spectroscopic analysis of the optical spectra 
               using FASTWIND]{Sim06}}
         \label{figsed}
   \end{figure}

The second important input to the photoionization model is 
the ionizing radiation field. We considered the stellar parameters 
derived for $\theta^1$\,Ori\,C by \cite{Sim06} and obtained the spectral 
energy distribution (SED) as predicted by the stellar atmopshere codes WM{\it basic} 
\citep{Pau01}, CMFGEN \citep{Hil98}, TLUSTY \citep{Hub95}, and FASTWIND \citep{Pul05}. 
Fig. \ref{figsed}
compares the four SEDs, also indicating the ionization energies for various
of the ions present in the Orion nebula. As commented in \cite{Sim08}, there 
are important discrepancies between the ionizing SEDs
predicted by the four stellar atmosphere codes. We considered the four SEDs
in our photoionization models, and it turned out that WM{\it basic} gives the best 
solution for this case (this argument is based on the simultaneous
fitting of the ionization degree constraints, corresponding to the ionization
of S$^+$, O$^+$, and Ar$^{2+}$; see also Figure \ref{figsed}). We also 
explored slight changes in the effective 
temperature, and found that the indicated stellar parameters (\Teff\,=\,39000 K, 
\grav\,=4.1) provide the best solution.

In all cases, a total number of hydrogen ionizing photons,  Q(H$^0$)=$10^{48.67}$ photons\,s$^{-1}$
was considered. This value was obtained assuming the same $m_v$ magnitude and extinction
as in \cite{Sim06}, but adopting a distance to the Orion nebula of 400 pc\footnote{This value agrees
better with recent determinations of the distance to the Orion nebula \citep[see][and 
references therein]{Men07}}, instead of the value originally adopted 450 pc.

%
\begin{table}[!t]
{\small
\begin{center}
\caption{\small List of nebular line ratios used to compare with the output of
the photoionization models and the corresponding observed values (from E04).
\label{tabmodel} }
\begin{tabular}{c|c|cc}
\noalign{\smallskip}
\tableline\tableline
Parameters         & Line ratio                                                 &  $O_{\rm obs}$    & e$_{\rm r}(O)$ \\
\tableline
\noalign{\smallskip}
\multicolumn{4}{l}{Density (constraint)} \\
\tableline
\Ne\nioni{Ar}{iv}   & \nioni{Ar}{iv}\,4740/4711                                  &  1.21   &  0.08  \\   
\Ne\nioni{O}{ii}   & \nioni{O}{ii}\,3726/3729                                    &  2.07   &  0.06  \\   
\Ne\nioni{S}{ii}  & \nioni{S}{ii}\,6731/6716                                     &  1.81   &  0.08  \\   
\tableline
\noalign{\smallskip}
\multicolumn{4}{l}{Ionization degree (constraint)} \\
\tableline											 
S$^{2+}$/S$^{+}$   & \nioni{S}{iii}\,9069/\nioni{S}{ii}\,6716+31  &  5.53     &  0.16 \\   
O$^{2+}$/O$^{+}$    & \nioni{O}{iii}\,5007/\nioni{O}{ii}\,3727         &  4.64   &  0.06   \\   
Ar$^{3+}$/Ar$^{2+}$  & \nioni{Ar}{iv}\,4711+40/\nioni{Ar}{iii}\,7135       &   0.0136  &   0.11 \\   
\tableline
\noalign{\smallskip}
\multicolumn{4}{l}{Abundances} \\								 
\tableline											 
O     & $100\times$ \nioni{O}{iii}\,5007/H$_{\beta}$               & 383	  &  0.03   \\
      & $100\times$ \ioni{O}{ii}\,4075/H$_{\beta}$  &    0.27   &   0.10  \\
C     & $100\times$ \ioni{C}{ii}\,4267/H$_{\beta}$                 &   0.23  &  0.05  \\
N    & $100\times$ \nioni{N}{ii}\,6584/H$_{\beta}$                &  37.7   &  0.05  \\
S     & $100\times$ \nioni{S}{iii}\,9069/H$_{\beta}$               &  30.2 	  & 0.14  \\
Ar   & $100\times$ \nioni{Ar}{iii}\,7135/H$_{\beta}$              &   16.2    &  0.07  \\
Ne  & $100\times$ \nioni{Ne}{iii}\,3869/H$_{\beta}$              &  22.9   &  0.04  \\
Fe    & $100\times$ \nioni{Fe}{iii}\,4659/H$_{\beta}$              &   0.55  &   0.05 \\
\tableline
\noalign{\smallskip}										 
\multicolumn{4}{l}{Temperature} \\								 
\tableline											 
\Te\nioni{O}{iii}   & \nioni{O}{iii}\,4363/5007                                  &  0.00339  &  0.03  \\
\Te\nioni{S}{iii}  & \nioni{S}{iii}\,6312/9069                                   &  0.0613   & 0.15  \\
\Te\nioni{O}{ii}  & \nioni{O}{ii}\,7325/3727                                     &  0.116    &   0.13  \\
\Te\nioni{N}{ii}   & \nioni{N}{ii}\,5755/6584                                    & 0.0180 &  0.06  \\
\tableline
\end{tabular}
\end{center}
}\end{table}

Finally, we took grains into account in the computation of all the models, assuming the 
``Orion type'' grain composition considered by Cloudy, scaled by a factor 1.1.
This scale factor was included to obtain the same abundance of Si, Mg, and Fe in grains
as computed in this paper (see Sect. \ref{sec-dust}).

To judge the quality of the model, we use the same procedure as \cite{Sta10}, plotting 
for each observable O (i.e. here for each relevant line ratio) the value of   
\begin {equation} 
  \kappa(O) = ({\rm log} O_{\rm mod} - {\rm log} O_{\rm obs})/\tau(O),
\end {equation} 
where $O_{\rm mod}$ is the value returned by the model, $O_{\rm obs}$ is the
observed value, and $\tau(O)$ the accepted tolerance in dex for this observable. For each observable, 
the value of  $\tau(O)$ is defined by   
\begin {equation} 
\tau(O) = {\rm log} (1+ e_{\rm r}(O)),
\end {equation}
where $e_{\rm r}(O)$ is the  maximum ``acceptable'' relative error on the observable.  A model 
is satisfactory only if for \textit{each} of the observables used as constraints, 
$\kappa(O)$ is found to be between $-1$ and $+1$. Table \ref{tabmodel} lists the values of the observables 
used as constraints to fit the model and the corresponding values of $e_{\rm r}(O)$ attributed 
to them (it also lists, for information, some other observables that are not used as constraints, such as line 
ratios indicative of the electron temperature or of the element abundances). 
 
\begin{figure*}[t!]
   \centering
   \includegraphics[width=3.5cm, angle=90]{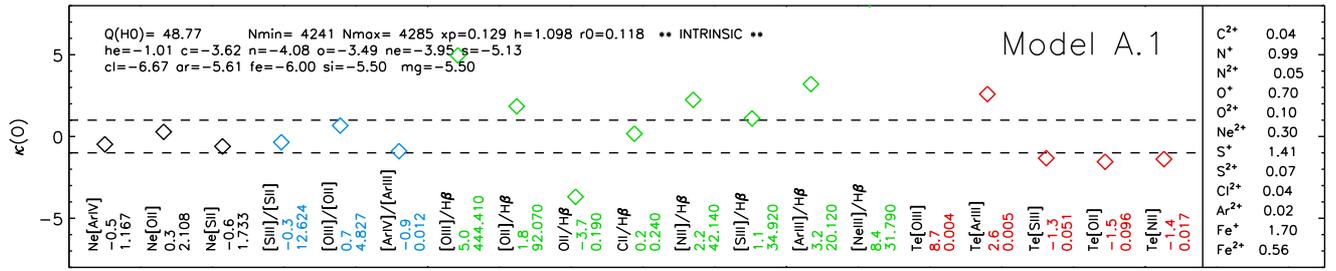}
      \caption{Results from a photoionization model in which the input abundances are the CEL ones given in  Table \ref{sum-ab}.
      The numbers to the right of each line ratio indicate the deviation from the 
      observed value in units of $\kappa(O)$ and the value obtained from the photoionization
      model (the observed value being given in Table \ref{tabmodel}). In the right 
      column are listed the logarithms of the icfs corresponding to this model.}
         \label{figmodelCEL}
   \end{figure*}

Figure \ref{figmodelCEL} is a graphic representation of one of our $best$ models and of the quality 
of the fits. In this model (Model B.1) the input abundances are the CEL ones given in Table 
\ref{sum-ab}.
In the right panel of the figure the icfs are listed corresponding to the model.\footnote{ In 
Tables \ref{table-nnears} and \ref{table-RLocne} are listed the values of icfs that were used 
to compute the element abundances. We estimated the uncertainties in the icfs by changing the 
input parameters of the photoionization model,  still keeping the values of $\kappa$ in the 
acceptable range.}
We may notice that using the CEL abundances implies a \Te\nioni{O}{iii} in the model much higher 
than  observed. This suggests that the input abundances (especially that of oxygen) might be too small.
Since we argue in Sect. \ref{discussion} that the gas phase oxygen abundance seems to be better estimated by 
RLs than by CELs, one may wonder whether it is correct to fit the ionization state of the nebula 
with ratios of CELs, as we have done. We believe that this procedure is reasonably correct, 
unless the discrepancy between RL and CEL abundances is much different in the low ionization and in 
the high ionization zone. 

\begin{figure*}[t!]
   \centering
   \includegraphics[width=3.5cm, angle=90]{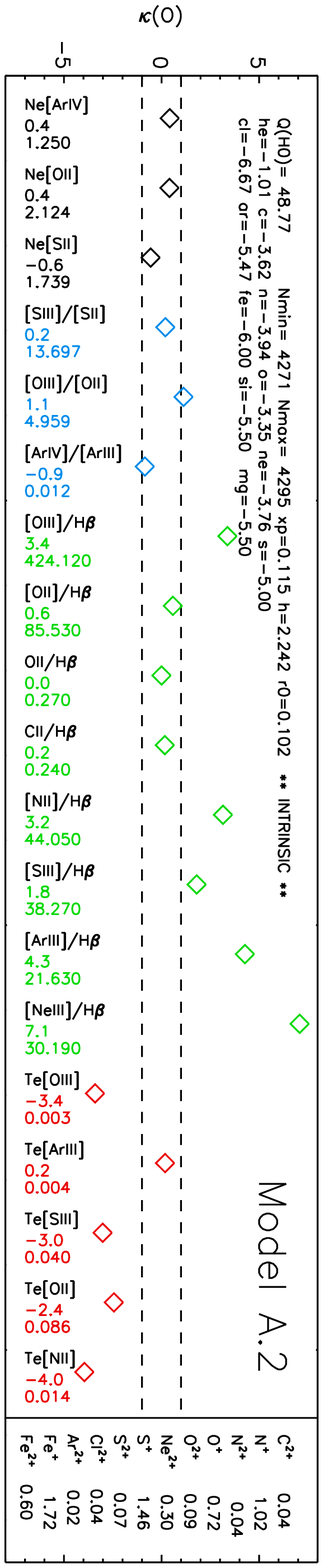}
      \caption{Results from a photoionization model in which the input abundances are the RL ones given in  Table \ref{sum-ab}.
      The numbers to the right of each line ratio indicate the deviation from the 
      observed value in units of $\kappa(O)$ and the value obtained from the photoionization
      model (the observed value being given in Table \ref{tabmodel}). In the right 
      column are listed the logarithms of the icfs corresponding to this model.}         
      \label{figmodelRL}
   \end{figure*}

Figure  \ref{figmodelRL} shows the results of another  model where this time the input 
abundances of C, O, and Ne are those obtained from the RL lines and the abundance of N is the 
CEL one increased by 0.14 dex.  The aim of this model is to discuss the energy budget problem as
 presented in Sect. \ref{discussion}.

\end{document}